\documentclass[12pt]{article}
\usepackage{a4wide}
\usepackage{latexsym}
\usepackage{amsmath}
\usepackage{amsfonts}
\usepackage{amscd}
\usepackage{cite}

\usepackage{pslatex}
\usepackage{graphicx}
\usepackage[latin1]{inputenc}
\usepackage[T1]{fontenc}

\def\bq{\begin{eqnarray}}
\def\eq{\end{eqnarray}}

\def\v{\verb}

\begin{document}

\thispagestyle{empty}

\begin{flushright}
  MZ-TH/11-22 
\end{flushright}

\vspace{1.5cm}

\begin{center}
  {\Large\bf The SISCone jet algorithm optimised for low particle multiplicities\\
  }
  \vspace{1cm}
  {\large Stefan Weinzierl\\
  \vspace{1cm}
      {\small \em Institut f{\"u}r Physik, Universit{\"a}t Mainz,}\\
      {\small \em D - 55099 Mainz, Germany}\\
  } 
\end{center}

\vspace{2cm}

\begin{abstract}\noindent
  {
The SISCone jet algorithm is a seedless infrared-safe cone jet algorithm.
There exists an implementation which is highly optimised for a large number of final state particles.
However, in fixed-order perturbative calculations with a small number of final state particles, it turns out that the 
computer time needed for the jet clustering of this implementation is comparable to the computer time of the matrix elements.
This article reports on an implementation of the SISCone algorithm optimised for low particle multiplicities. 
   }
\end{abstract}

\vspace*{\fill}

\newpage 

{\bf\large PROGRAM SUMMARY}
\vspace{4mm}
\begin{sloppypar}
\noindent   {\em Title of program\/}: \v/siscone_parton/ \\[2mm]
   {\em Version\/}: 1.0.1 \\[2mm]
   {\em Catalogue number\/}: \\[2mm]
   {\em Program obtained from\/}: {\tt http://wwwthep.physik.uni-mainz.de/\~{}stefanw/software.html} \\[2mm]
   {\em E-mail\/}: {\tt stefanw@thep.physik.uni-mainz.de}\\[2mm]
   {\em License\/}: GNU Public License \\[2mm]
   {\em Computers\/}: all \\[2mm]
   {\em Operating system\/}: all \\[2mm]
   {\em Program language\/}: {\tt C++     } \\[2mm]
   {\em Memory required to execute\/}: 
         Negligible for low parton multiplicities. \\[2mm]
   {\em Other programs called\/}: none \\[2mm]
   {\em External files needed\/}: none \\[2mm]
   {\em Keywords\/}:  Jet algorithms.\\[2mm]
   {\em Nature of the physical problem\/}: 
         Clustering of particles into jets. \\[2mm]
   {\em Method of solution\/}: 
         Infrared-safe cone algorithm.\\[2mm] 
   {\em Restrictions on complexity of the problem\/}: 
         Hard-coded restriction to 64 final state particles on 64-bit machines, 
         but recommended to be used only for configurations with up to 10 final state particles. \\[2mm]
   {\em Typical running time\/}:
         Depending on the number of final state particles, ${\cal O}(\mathrm{\mu s})$ for configurations up to 5 final state particles.
\end{sloppypar}

\newpage

\section{Introduction}
\label{sec:intro}

The clustering of particles into hadronic jets is relevant to all current high-energy collider experiments.
The clustering procedure is described by a jet algorithm.
The available jet algorithms can be divided into two classes: Sequential recombination algorithms and cone-type algorithms.
For a comparison between experiment and theory it is essential that a jet algorithm is infrared-safe.
Infrared-safe observables can be calculated reliably within perturbation theory.
The cone-type algorithms have suffered for a long time from not being infrared-safe.
This problem has been solved with the invention of the SISCone jet algorithm, 
which is a seedless infrared-safe cone jet algorithm \cite{Kidonakis:1998bk,Blazey:2000qt,Salam:2007xv}.

For a correct comparison between experiment and theory one has to use exactly the same jet algorithm in both cases.
However, the situations in which the jet algorithm needs to be used are quite different:
Experimentalists need to cluster of the order of one hundred particles, while in higher-order calculations within fixed-order
perturbation theory one usually deals only with a few partons in the range from two to five or six.
The original proposal of the SISCone algorithm in \cite{Blazey:2000qt} described a method, which takes ${\cal O}(n \cdot 2^n)$ time
to cluster $n$ particles.
This is impractical in the experimental case.
A major breakthrough was therefore the invention of a method, which performs this task in ${\cal O}(n^2 \ln n)$ time \cite{Salam:2007xv}.
In addition, the authors of ref.~\cite{Salam:2007xv} clarified and fixed several ambiguities related to residual infrared unsafety.
The implementation by the authors of ref.~\cite{Salam:2007xv} has become the reference implementation of the SISCone algorithm.
This method together with the corresponding implementation was mainly responsible for the fact,
that experimentalists nowadays use frequently the infrared-safe SISCone jet algorithm.

Let us now turn to the theoretical side:
Next-to-leading order (NLO) or next-to-next-to-leading order (NNLO) calculations are usually done with the help of the subtraction method.
This introduces a collection of subtraction terms for the real emission part.
Each subtraction term has its own configuration of parton momenta, in general different from the other ones and different from the configuration 
of the matrix element it approximates.
For jet observables this implies that the jet algorithm has to be run once for each subtraction term.
For sequential recombination algorithms like the $k_\perp$- or anti-$k_\perp$-algorithms \cite{Stirling:1991ds,Catani:1993hr,Ellis:1993tq,Cacciari:2005hq,Cacciari:2008gp} this is not a problem, since the CPU time
for the jet algorithm is negligible against the CPU time required for the matrix element associated with the subtraction term.
However, in a recent comparison of different jet algorithms at NNLO in electron-positron annihilation \cite{Weinzierl:2010cw}
it turned out that the CPU time of the reference implementation of the SISCone algorithm was at the order of the CPU time of the matrix
elements.
This slowed down the computation of the NNLO prediction for the SISCone jet rates significantly compared to the NNLO predictions
for the jet rates with other jet algorithms, in particular when results for different cone sizes were calculated in parallel.
(In ref.~\cite{Weinzierl:2010cw} the results for 100 different cone sizes where calculated in parallel.)
To fix this problem, a dedicated implementation of the SISCone algorithm optimised for low particle multiplicities has been written.
These routines can be of use for any calculation within fixed-order perturbation theory using the SISCone algorithm.

In this article I report on the implementation of the SISCone algorithm optimised for low particle multiplicities. 
The implementation includes a version for hadronic collisions as well as a version for electron-positron annihilation.
The implementation is written in C++. Wherever possible I kept the same syntax as the reference implementation of \cite{Salam:2007xv}.
Efficiency issues require a few modifications of the syntax, which are all documented in this article.

It should be clearly stated the implementation of this article is targeted for the case of a few final state particles.
In this case the implementation of this article can give a significant performance boost, up to a factor $70$ 
for two-particle configurations.
For final states where the number of particles is at the order of one hundred, the implementation of \cite{Salam:2007xv} should be used.
The cross-over occurs around $10$ final state particles.

This paper is organised as follows: 
In the next section I review the SISCone jet algorithm.
Section~\ref{sect:implementation} describes the optimisation techniques used in the low parton multiplicity case.
Section~\ref{sect:howto} gives detailed information on how to use the program.
The performance is reported in section~\ref{sect:performance}.
Finally, section~\ref{sect:conclusions} contains a summary and the conclusions.


\section{The SISCone jet algorithm}
\label{sect:definition}

In this section I review the definition of the SISCone jet algorithm \cite{Salam:2007xv}.
There are two variants of the SISCone jet algorithm, one variant is adapted to hadronic collisions, while the other
variant is adapted to electron-positron collisions.
The hadronic version uses the rapidity $y$ and the azimuthal angle $\phi$ as the basic variables and corresponds to 
a cylindrical geometry.
The version for electron-positron annihilation uses instead the angle between particles as the basic variable
and corresponds to a spherical geometry.
I will first describe the common features of both versions and give the specific features for each version in the end.

The SISCone jet algorithm depends on the four parameters $R$, $f$, $n_{\mathrm{pass}}$ and $v_{\mathrm{min}}$.
The most important parameters are the cone size $R$ and the overlap parameter $f$.
The parameter $n_{\mathrm{pass}}$ specifies how often the procedure for finding stable cones is maximally iterated. This parameter can be set to infinity.
The parameter $v_{\mathrm{min}}$ defines the minimal value of the quantity $v_{\mathrm{threshold}}$ (to be defined below) required for a jet.
In addition to these four numbers we have to specify a distance measure $R(i,J)$ between a particle $i$ and a cone axis corresponding 
to the protojet $J$ and three functions
$v_{\mathrm{threshold}}(J)$, $v_{\mathrm{ordering}}(J)$ and $v_{\mathrm{overlap}}(J)$ related to the split merge procedure.
The two versions for hadronic collisions and electron-positron annihilation will only differ in the definition of the distance measure
and in the definition of the functions $v_{\mathrm{threshold}}$, $v_{\mathrm{ordering}}$ and $v_{\mathrm{overlap}}$.

The SISCone jet algorithm is specified as follows:
\begin{enumerate}
\item Put the set of current particles equal to the set of all particles in the event and set $i_{pass}=0$.
\item For the current set of particles find all stable cones with cone size $R$. 
\item Each stable cone is added to the list of protojets.
\item Remove all particles that are in stable cones from the list of current particles and increment $i_{pass}$.
\item If $i_{pass}<n_{pass}$ and some new stable cones have been found in this pass, go back to step 2.
\item Run the split-merge procedure with overlap parameter $f$ and threshold $v_{\mathrm{min}}$.
\end{enumerate}
A set of particles defines a cone axis, which is given as the sum of the momenta of all particles in the set.
A cone is called stable for the cone size $R$, 
if all particles defining the cone axis have a distance measure smaller than $R$ to the cone
axis and if all particles not belonging to the cone have a distance measure larger than $R$ with respect to the cone axis.
The four-momentum of a protojet is the sum of the four-momenta of the particles in the protojet.
This corresponds to the E-scheme.
Two protojets are called overlapping, if they share at least one particle.

Let us now turn to the split-merge procedure:
The split-merge procedure depends on the two parameters $v_{\mathrm{min}}$ and $f$, and on the three functions 
$v_{\mathrm{threshold}}$, $v_{\mathrm{ordering}}$ and $v_{\mathrm{overlap}}$.
The split-merge procedure discards all protojets with $v_{\mathrm{threshold}} < v_{\mathrm{min}}$, therefore $v_{\mathrm{min}}$ is a threshold value
a jet must have.
In addition, the split-merge procedure requires an infrared-safe ordering variable for the protojets. 
This variable is denoted $v_{\mathrm{ordering}}$.
Furthermore, if two protojets overlap, we need a quantity which describes the overlap. This quantity is denoted $v_{\mathrm{overlap}}$.
The split-merge procedure is defined as follows:
\begin{enumerate}
\item Remove all protojets with $v_{\mathrm{threshold}} < v_{\mathrm{min}}$. 
\item Find the protojet $I$ with the highest value of $v_{\mathrm{ordering}}$.
\item Among the remaining protojets find the one ($J$) with highest value $v_{\mathrm{ordering}}$ that overlaps with $J$.
\item If there is such an overlapping jet then compute the quantities $v_{\mathrm{overlap}}(I \cap J)$ and $v_{\mathrm{overlap}}(J)$.
\begin{enumerate}
\item If $v_{\mathrm{overlap}}(I \cap J) < f v_{\mathrm{overlap}}(J)$ assign each particle that is shared between the two protojets to the protojet whose
axis is closest. Recalculate the momenta of the protojets.
\item If $v_{\mathrm{overlap}}(I \cap J) \ge f v_{\mathrm{overlap}}(J)$ merge the two protojets into a single new protojet and remove the two original
ones.
\end{enumerate}
\item Otherwise, if no overlapping jet exists, then add $I$ to the list of jets and remove it from the list
of protojets.
\item As long as there are protojets left, go back to step 1.
\end{enumerate}
It remains to specify the distance measure $R(i,J)$
and the three functions $v_{\mathrm{threshold}}(J)$, $v_{\mathrm{ordering}}(J)$ and $v_{\mathrm{overlap}}(J)$.
These quantities differ in the two versions of the SISCone algorithm.
We have
\begin{align}
 &
   & & \mbox{cylindrical}
   & & \mbox{spherical}
 \nonumber \\
 & R(i,J): 
    & & d_{i,J} = \sqrt{ \left(y_i-y_J\right)^2 + \left( \phi_i -\phi_J \right)^2} 
    & & \theta_{i,J} = \arccos \frac{\vec{p}_i \cdot \vec{p}_J}{\left|\vec{p}_i \right| \left| \vec{p}_J\right|}
 \nonumber \\
 & v_{\mathrm{threshold}}(J):
    & & p_{\perp}(J) = \left| \sum\limits_{k \in J} \vec{p}_{\perp,k} \right|
    & & E(J) = \sum\limits_{k \in J} E_k
 \nonumber \\
 & v_{\mathrm{ordering}}(J):
    & & \tilde{p}_{\perp}(J) = \sum\limits_{k \in J} \left| \vec{p}_{\perp,k} \right|
    & & \tilde{E}(J) = \sum\limits_{k \in J} E_k \left( 1 + \frac{\left|\vec{p}_J\right|^2}{E_J^2} \sin^2 \theta_{k,J} \right)
 \nonumber \\
 & v_{\mathrm{overlap}}(J):
    & & \tilde{p}_{\perp}(J) = \sum\limits_{k \in J} \left| \vec{p}_{\perp,k} \right|
    & & E(J) = \sum\limits_{k \in J} E_k 
 \nonumber
\end{align}
In this table the indices $i$ and $k$ refer to a particle, while the index $J$ refers to the four-vector of a protojet.
In the cylindrical version (which is appropriate for hadronic collisions) the distance measure is calculated from the rapidity $y$ and
the azimuthal angle $\phi$.
The threshold function $v_{\mathrm{threshold}}(J)$ is given as the absolute value of the sum of the transverse momenta $p_{\perp,k}$ of the particles $k$
making up the protojet $J$.
The ordering function $v_{\mathrm{ordering}}(J)$ and the overlap function $v_{\mathrm{overlap}}(J)$ are given as the sum
of the absolute values of the transverse momenta $p_{\perp,k}$ of the particles $k$
making up the protojet $J$.
Since the main application of this implementation is within fixed-order perturbation theory infrared-safeness is essential.
Therefore choices of the ordering function which are not infrared-safe are not considered.
The choice for $v_{\mathrm{ordering}}(J)$ given above is the recommended one of ref.~\cite{Salam:2007xv}.

In the spherical version (which is appropriate for electron-positron annihilation) the distance measure is given by the angle between the three-momentum
$\vec{p}_i$ of particle $i$ and the three-momentum $\vec{p}_J$ of the protojet axis of the protojet $J$.
The threshold function $v_{\mathrm{threshold}}(J)$ and the overlap function $v_{\mathrm{overlap}}(J)$ are given by the sum of the energies of the 
particles $k$ making up the protojet $J$.
The ordering function $v_{\mathrm{ordering}}(J)$ is given by the quantity 
\bq
 \tilde{E}(J) & = & \sum\limits_{k \in J} E_k \left( 1 + \frac{\left|\vec{p}_J\right|^2}{E_J^2} \sin^2 \theta_{k,J} \right).
\eq
Note that this is the function, which is actually implemented in the code of \cite{Salam:2007xv},
and not the function
\bq
 \sum\limits_{k \in J} E_k \left( 1 + \sin^2 \theta_{k,J} \right).
\eq


\section{The implementation for low particle multiplicities}
\label{sect:implementation}

The standard implementation \cite{Salam:2007xv} of the SISCone algorithm 
is highly optimised for the experimental situation, where one faces the challenge to cluster
of the order one hundred particles.
In this situation the asymptotic run-time behaviour of the algorithm is very important. 
The asymptotic run-time behaviour is dominated by the time needed to find all stable cones
and is ${\cal O}(n^2 \ln n)$ for $n$ final state particles in the implementation of \cite{Salam:2007xv}.
The CPU time of the parts of the algorithm which do not grow so fast are less important if one deals with the
order of one hundred particles.

However, the situation is different if only a few final state particles need to be clustered.
In this situation the asymptotic run-time behaviour is less important, and all parts of the algorithm have to be taken
into account for the optimisation. 
It turns out that it is most efficient to use for the low particle multiplicity case a method 
which reduces significantly the computational cost of the parts which would be sub-dominant in the asymptotic regime.
The implementation of this paper uses the method of \cite{Blazey:2000qt} to find all stable cones. This method
grows like ${\cal O}(n \cdot 2^n)$ for $n$ final state particles.
This is acceptable since the intention is to use this implementation only in the case of small values of $n$.

The implementation is written in C++. There are a few standard optimisation techniques which are used to make the code fast:
First of all the intention is that this implementation is used within theoretical fixed-order calculations.
The authors of fixed-order numerical programs usually have their own implementation of a four-vector class, which stores the momenta
of the final state particles.
Copying the data to a similar class which is used by the jet algorithm can take a significant amount of CPU time.
This can be avoided by making the implementation of the SISCone jet algorithm a template.
The implementation of this paper works with any four-vector class, under the condition that the four-vector class
provides a few standard methods.
A simple four-vector class which has all the required methods comes with the implementation.

The input to the jet algorithm is a list (actually a {\tt std::vector}) of $n$ four-vectors, representing the $n$ final-state particles.
After clustering has been done, the content of a jet is represented by an unsigned integer. 
Within this unsigned integer, the $j$-th bit corresponds to particle $j$.
A value of $1$ of the corresponding bit indicates that the particle belongs to the jet, while a value
of $0$ means that the particle does not belong to the jet.
In this way one can represent up to $64$ particles in an unsigned integer on $64$-bit machines
(and up to $32$ particles in an unsigned integer on $32$-bit machines).
However, these numbers should be taken as a hard-coded upper limit, beyond which it is not possible to use this implementation.
In practise one would like to switch already earlier on to the implementation of \cite{Salam:2007xv} with the better
asymptotic run-time behaviour.
The advantage of using unsigned integers to represent jets is given by the fact that many operations (like merging) can be carried
out through fast bit-operations.

The implementation of the spherical version of the SISCone jet algorithm uses internally the distance measure
\bq
 y_{i,J} & = & 1 - \cos \theta_{i,J}
\eq
instead of $\theta_{i,J}$. $y_{i,J}$ is a monotonic function of $\theta_{i,J}$ in the interval $[0,\pi]$.
Using $y_{i,J}$ has the advantage that the distance can be calculated without any call to a trigonometric function.
The value of $y_{i,J}$ is calculated as
\bq
 y_{i,J} & = & 1 - \frac{\vec{p}_i}{\left|\vec{p}_i\right|} \cdot \frac{\vec{p}_J}{\left|\vec{p}_J\right|}.
\eq
This requires only a call to the \v/sqrt/-function for the normalisation of the spatial three-vectors $\vec{p}$.
All normalised spatial three-vectors are pre-computed at the start of the jet algorithm and stored in an array.

In the cylindrical case the distance measure involves the rapidity $y$ and the azimuthal angle $\phi$.
This involves a call to the \v/log/-function and a call to the \v/atan2/-function. 
These calls are rather expensive and lead to the effect that the cylindrical version requires more computer time.
In order to avoid some of these function calls, the following optimisation is implemented in the program:
At the start of the jet algorithm, all azimuthal angles are pre-computed and stored in an array, but the rapidities
are not yet computed.
In the search for stable cones some cones can immediately be discarded based
on the information of the azimuthal angle alone.
These are the cones $J$, which have a particle $i$ belonging to the cone $J$, whose distance in azimuthal angle is already 
larger than the cone size $R$:
\bq
 \tilde{d}_{i,J}^2 & = & \left( \phi_i - \phi_J \right)^2 > R^2.
\eq
These cones are never stable, since adding the rapidity distance would only increase the distance measure.
For those cones the calculation of the rapidity can be avoided.

A second optimisation technique can be applied to events, which have a total $p_\perp$-sum of zero.
This is the case for pure QCD events, but not for mixed QCD-electroweak events like $W + \mbox{jets}$.
If the total $p_\perp$-sum is zero,
only half of the values of the azimuthal angle (and of the transverse momenta) need to be computed.
The other half can be obtained from transverse momentum conservation without a call to the \v/atan2/-function.
In the implementation this additional optimisation can be turned on by setting a flag that the total $p_\perp$-sum is zero.


\section{How to use the program}
\label{sect:howto}

The implementation of the SISCone jet algorithm is written in C++ and can be obtained from
\\
\\
{\tt http://wwwthep.physik.uni-mainz.de/\~{}stefanw/software.html}\\
\\
After unpacking, the directory will contain the four files
\begin{verbatim}
siscone_parton.h  fourvector.h  example.cc  example_spherical.cc
\end{verbatim}
The file \v/siscone_parton.h/ contains the implementation of the SISCone jet algorithm,
defined within the namespace \v/siscone_jet_algorithm/.
The SISCone jet algorithm is implemented as a template class, therefore all code is contained in the header file
\v/siscone_parton.h/.
No compilation or installation is required, the implementation can be used by including the
header file in the source code of the user.
The version of the SISCone jet algorithm which is based on the cylindrical geometry and which is appropriate for hadronic
collisions is implemented in
\begin{verbatim}
template<class T> class siscone_parton
\end{verbatim}
The version based on the spherical geometry, which is appropriate for electron-positron annihilation is implemented in
\begin{verbatim}
template<class T> class siscone_spherical_parton
\end{verbatim}
In both cases the template argument is a class implementing four-vectors and denoted by \v/T/ in the declarations above.
The four-vector class must have a few standard methods, which are summarised in the following fragment of an example header file:
\begin{verbatim}
class fourvector {

 public :  
  fourvector(void);

  fourvector & sum_up(const fourvector & q);
  fourvector & rescale(double c);

  double transverse_momentum(void) const;
  double rapidity(void) const;
  double azimuthal_angle(void) const;

  double spatial_norm(void) const;
  double spatial_scalar_product(const fourvector & q) const;
  double energy_component(void) const;
};
\end{verbatim}
In detail, the four-vector class has to provide a default constructor, which initialises the four components of the four-vector with zero.
The method \v/sum_up(q)/ adds the four-vector \v/q/ to the current four-vector.
The method \v/rescale(c)/ multiplies all components of the four-vector by the factor \v/c/.
The default constructor and the two methods \v/sum_up/ and \v/rescale/ are required by both versions of the SISCone jet algorithm.

The class \v/siscone_parton/ which uses the cylindrical geometry requires in addition the following three methods:
A method \v/transverse_momentum()/ which returns the transverse momentum with respect to the beam axis.
With the beam-axis along the $z$-axis the transverse momentum is given by 
\bq
 p_\perp & = & \sqrt{p_x^2+p_y^2},
\eq
where $p_x$ and $p_y$ denotes the $x$- and the $y$-component of the four-vector $p$, respectively.
Secondly, a method \v/rapidity()/ which returns the rapidity of the four-momentum. With the beam along the $z$-axis, the rapidity is given 
by 
\bq
 y  & = & \frac{1}{2} \ln \frac{p_t+p_z}{p_t-p_z},
\eq
where $p_t$ denotes the energy component and $p_z$ denotes the $z$-component of the four-vector $p$.
Finally, a method \v/azimuthal_angle()/ which returns the azimuthal angle $\phi$ of the four-vector with respect to the beam axis. This angle
obeys the relation
\bq
 \tan \phi & = & \frac{p_y}{p_x},
\eq
if the beam-axis is again along the $z$-axis.
The three methods \v/transverse_momentum/, \v/rapidity/ and \v/azimuthal_angle/ are not needed if just the spherical version of the 
SISCone jet algorithm is used.

The class \v/siscone_spherical_parton/ which uses the spherical geometry requires in addition the following three methods:
A method \v/spatial_norm()/ which returns the norm of the three spatial components of the four-vector:
\bq
 \left| \vec{p} \right|^2 & = & p_x^2 + p_y^2 + p_z^2.
\eq
Secondly, a method \v/spatial_scalar_product(q)/ which returns the scalar product of the spatial components with the ones of the four-vector \v/q/:
\bq
 \vec{p} \cdot \vec{q} & = & p_x q_x + p_y q_y + p_z q_z.
\eq
Finally, a method \v/energy_component()/, which returns the energy component $p_t$ of the four-vector.
The three methods \v/spatial_norm/, \v/spatial_scalar_product/ and \v/energy_component/ 
are not needed if just the cylindrical version of the 
SISCone jet algorithm is used.

The file \v/fourvector.h/ gives an example of a four-vector class which has all the required methods.
Since the required methods are rather simple, they can be implemented as inline functions.
Therefore all code relevant to the four-vector class is contained in the header file \v/fourvector.h/.

The SISCone jet algorithm based on the cylindrical geometry is initialised by
\begin{verbatim}
siscone_parton<fourvector> jet_finder(n_max);
\end{verbatim}
The template argument \v/fourvector/ should be replaced by the name of the appropriate four-vector class.
The variable \v/n_max/ gives the maximal number of final state particles.
In a fixed-order perturbative calculation \v/n_max/ is usually set by the number of final state particles in the real emission contribution.
The constructor allocates the memory which is needed for the jet algorithm.
Since the CPU time for the allocation and deallocation of memory is not negligible, it is recommended that within a Monte Carlo integration
the constructor is called exactly once.

The spherical version of the SISCone jet algorithm is correspondingly initialised by
\begin{verbatim}
siscone_spherical_parton<fourvector> jet_finder_spherical(n_max);
\end{verbatim}

In both versions the jet algorithm is invoked by a call to the method
\begin{verbatim}
int compute_jets(const std::vector<fourvector> & particles, double R, 
                 double f, int n_pass_max=0, double v_min=0.0);
\end{verbatim}
The input parameters of the method are: a list of four-vectors of the final state particles, given as a \v/std::vector<fourvector>/.
The input parameters \v/R/ and \v/f/ denote the cone size and the overlap parameter, respectively.
The maximum number of passes is given by \v/n_pass_max/.
A zero value for \v/n_pass_max/ is treated as infinity.
This is the default choice for \v/n_pass_max/.
The minimal value of the threshold parameter is given by \v/v_min/. In the cylindrical version this is the $p_\perp$-value, in the spherical version
it is the energy $E$.
The default choice for this parameter is zero.
The name and the arguments of the method \v/compute_jets/ are identical to the ones in the implementation of \cite{Salam:2007xv}.
This should make it easy to switch between the two implementations and to use each implementation where it performs best.

The method \v/compute_jets/ returns the number of jets $n_{\mathrm{jet}}$.
In addition, the content of the jets can be found in the data member
\begin{verbatim}
template<class T> class siscone_parton
{
 public:
  std::vector<unsigned> final_jets;
};
\end{verbatim}
The class \v/siscone_spherical_parton/ has also the data member \v/final_jets/.
After a call to \v/compute_jets/ this vector contains $n_{\mathrm{jet}}$ elements. Each entry is an unsigned integer.
If the $j$-th bit of entry $i$ is set, it implies that particle $j$ belongs to jet $i$.

The class \v/siscone_parton/ has in addition a method
\begin{verbatim}
 public:
  void set_flag_total_pt_is_zero(bool flag);
\end{verbatim}
For events with a total $p_\perp$-sum of zero, this flag can be set to \v/true/.
The jet algorithm will then use special optimisations, which are valid if the total $p_\perp$-sum is zero.
This can be used for pure QCD events, but should not be used for mixed QCD+electroweak events like $W+\mbox{jets}$-events.
The default choice is \v/false/.

The file \v/example.cc/ contains a simple example program, which shows how to use the jet algorithm.
The listing of the test program is as follows:
\begin{verbatim}
#include <iostream>
#include <vector>

#include "fourvector.h"
#include "siscone_parton.h"

int main()
{
  using namespace siscone_jet_algorithm;

  // intialise the SISCone algorithm for maximally 4 final state particles
  int n_max = 4;
  siscone_parton<fourvector> jet_finder(n_max);

  // define the parameters for the SISCone algorithm
  double R      = 0.9;
  double f      = 0.5;
  int n_pass    = 0;
  double pt_min = 5.0;

  // some final-state momenta
  std::vector<fourvector> particles(4);
  particles[0] = fourvector(1.87116,-1.08275,1.33996,-0.730343);
  particles[1] = fourvector(33.3795,1.00171,28.1084,-17.9751);
  particles[2] = fourvector(32.4375,-12.8963,-13.154,26.6992);
  particles[3] = fourvector(22.3118,12.9774,-16.2944,-7.99375);

  // call the jet algorithm
  int n_jet = jet_finder.compute_jets( particles, R, f, n_pass, pt_min );

  // output
  std::cout << "Number of jets = " << n_jet << std::endl;
  for (int j=0; j<n_jet; j++) {
    std::cout << " Particle content of jet " << j << " : ";
    for (int i=0; i<n_max; i++) {
      if ( jet_finder.final_jets[j] & (1<<i) ) std::cout << i << " ";
    }
    std::cout << std::endl;
  }

  return 0;
}
\end{verbatim}
The program initialises first the jet algorithm for maximally four final state particles.
It then defines the parameters for the jet algorithm and a set of four final state momenta.
The call to the method \v/compute_jets/ invokes the jet algorithm.
After the jets have been computed, the program outputs the number of jets as well as the particle content of each jet.
The program can be compiled with the command
\begin{verbatim}
g++ -o example example.cc
\end{verbatim}
Running the program will produce the output
\begin{verbatim}
Number of jets = 3
 Particle content of jet 0 : 0 1 
 Particle content of jet 1 : 3 
 Particle content of jet 2 : 2 
\end{verbatim}
We see that three jets have been found. Particles $0$ and $1$ have been clustered together to form jet $0$.
Jet $1$ consists of the particle $3$, while jet $2$ consists of particle $2$.

The file \v/example_spherical.cc/ is very similar to the file \v/example.cc/, but uses the spherical version of the SISCone
jet algorithm instead.
The example for the spherical version uses the same set of momenta of the final state particles.
Running the program \v/example_spherical/ will produce the output
\begin{verbatim}
Number of jets = 3
 Particle content of jet 0 : 0 1 
 Particle content of jet 1 : 2 
 Particle content of jet 2 : 3  
\end{verbatim}


\section{Checks and performance}
\label{sect:performance}

In this section I compare the implementation of the SISCone jet algorithm of this article with the implementation
of \cite{Salam:2007xv} as a function of the number $n$ of final state particles.
The number $n$ of final state particles ranges in the comparison between $2$ and $11$.
For each $n$ I generate ${\cal O}(10^6)$ random events and I cluster the final state particles with the two implementations of
the SISCone jet algorithm.
The two implementations find exactly the same number of jets for all events.
Let us first consider in more detail the clustering based on the cylindrical geometry appropriate to hadronic collisions.
\begin{figure}
\begin{center}
\includegraphics[bb= 125 460 490 710,width=0.8\textwidth]{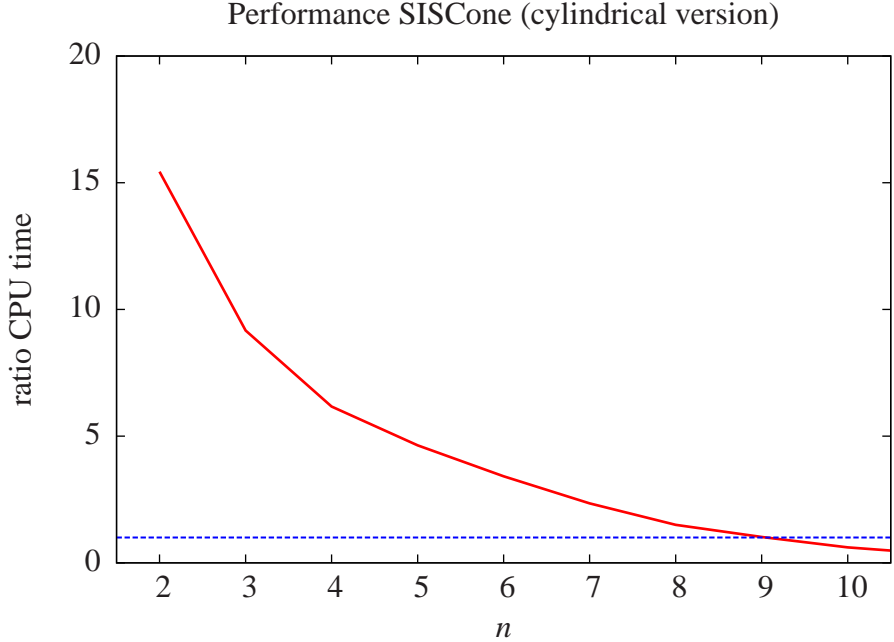}
\end{center}
\caption{
Speed-up factor as a function of the number $n$ of final state particles
for the cylindrical version of the SISCone jet algorithm.
Shown is the ratio of the CPU time of the implementation of \cite{Salam:2007xv} by the CPU time of this implementation.
The blue line shows a ratio of $1$. 
The cross-over occurs between $9$ and $10$ final state particles.
}
\label{fig_performance}
\end{figure}
Fig.~\ref{fig_performance} shows the speed-up factor 
as a function of the number $n$ of final state particles.
In this figure the ratio of the CPU time of the implementation of \cite{Salam:2007xv} by the CPU time of this implementation
is shown.
\begin{table}
\begin{center}
\begin{tabular}{|c|c|c|c|c|c|c|c|c|c|c|}
\hline
 $n$ & $2$ & $3$ & $4$ & $5$ & $6$ & $7$ & $8$ & $9$ & $10$ \\
 \hline
 this work & $0.3$ & $1.1$ & $2.4$ & $4.4$ & $8.0$ & $15$ & $28$ & $51$ & $100$ \\
 ref.~\cite{Salam:2007xv} & $4.9$ & $10$ & $15$ & $20$ & $27$ & $36$ & $42$ & $52$ & $61$ \\
\hline
\end{tabular}
\caption{\label{table_performance}
Average CPU time in $\mu\mbox{s}$ for the clustering of $n$ particles with the cylindrical version of the SISCone algorithm on a standard PC.
}
\end{center}
\end{table}
The absolute timings of the two implementations are shown in table~\ref{table_performance}. 
It can be seen that the implementation of this article is faster up to nine final state particles.
A speed-up factor of $15$ is obtained for configurations with two final state particles.
This factor is for example relevant to a forthcoming NNLO calculation of inclusive single jet production in $p p$ collisions.
In this calculation most subtraction terms have configurations with two final state particles.
\begin{figure}
\begin{center}
\includegraphics[bb= 125 460 490 710,width=0.8\textwidth]{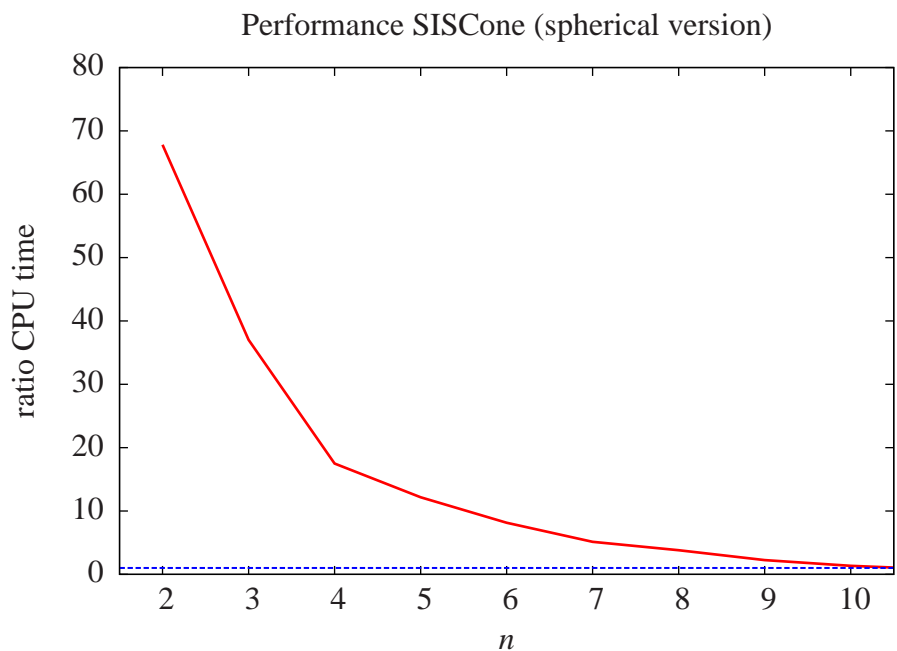}
\end{center}
\caption{
Speed-up factor as a function of the number $n$ of final state particles
for the spherical version of the SISCone jet algorithm.
Shown is the ratio of the CPU time of the implementation of \cite{Salam:2007xv} by the CPU time of this implementation.
The blue line shows a ratio of $1$. 
The cross-over occurs between $10$ and $11$ final state particles.
}
\label{fig_performance_spherical}
\end{figure}
Let us now consider the spherical version of the SISCone jet algorithm.
Fig.~\ref{fig_performance_spherical} shows the corresponding plot for the spherical geometry, appropriate
to electron-positron annihilation.
The absolute timings of the two implementations are shown in table~\ref{table_performance_spherical}. 
\begin{table}
\begin{center}
\begin{tabular}{|c|c|c|c|c|c|c|c|c|c|c|c|}
\hline
 $n$ & $2$ & $3$ & $4$ & $5$ & $6$ & $7$ & $8$ & $9$ & $10$ & $11$ \\
\hline
 this work & $0.1$ & $0.4$ & $1.1$ & $2.4$ & $4.7$ & $9.3$ & $16$ & $33$ & $65$ & $130$ \\
 ref.~\cite{Salam:2007xv} & $6.7$ & $13$ & $19$ & $29$ & $38$ & $48$ & $61$ & $74$ & $88$ & $105$ \\
\hline
\end{tabular}
\caption{\label{table_performance_spherical}
Average CPU time in $\mu\mbox{s}$ for the clustering of $n$ particles with the spherical version of the SISCone algorithm on a standard PC.
}
\end{center}
\end{table}
Here, the speed-up factor is almost $70$ for two-particle configurations and roughly $35$ for three-particle configurations.
The implementation of this article is faster up to $10$ particles.
The speed-up factor of $35$ for three-particle configurations was essential in the recent NNLO calculation
of three-jet rates in electron-positron annihilation \cite{Weinzierl:2010cw}.
Without this speed improvement the CPU time for the jet clustering would have out-weighted the CPU time of the corresponding matrix elements.

One observes that the speed-up factor is higher in the spherical version than in the cylindrical version.
This can be understood by the following two facts:
First, as explained in section~\ref{sect:implementation} the spherical version can be coded in such a way that 
no trigonometric function needs to be evaluated in the main part of the algorithm.
On the other hand, the cylindrical version requires the evaluation of \v/log/- and \v/atan2/-functions.
It is therefore expected that the implementation of the cylindrical version requires more CPU time compared to the
spherical version. In the implementation of this article this is indeed the case.
The second reason is rather surprising: It turns out that in the implementation of \cite{Salam:2007xv} the
cylindrical version performs better than the spherical version.


\section{Conclusions}
\label{sect:conclusions}

In this article I reported on an implementation of the SISCone jet algorithm, which is optimised for low particle multiplicities.
This implementation can be used for theoretical calculations in fixed-order perturbation theory, where the number of final state
particles is rather low. 
The implementation of this article performs better then the reference implementation of \cite{Salam:2007xv}
up to roughly nine final state particles.
For a higher number of final state particles the reference implementation of \cite{Salam:2007xv} should be used.
For a low number of final state particles the implementation of this article can lead to a speed-up factor up to $70$.

\subsection*{Acknowledgements}

I would like to thank G. Salam and G. Soyez for useful discussions.

\bibliography{/home/stefanw/notes/biblio}

\begin{thebibliography}{1}

\bibitem{Kidonakis:1998bk}
N.~Kidonakis, G.~Oderda, and G.~F. Sterman,
\newblock Nucl. Phys. {\bf B525}, 299 (1998), arXiv:hep-ph/9801268.

\bibitem{Blazey:2000qt}
G.~C. Blazey {\em et~al.},
\newblock (2000), arXiv:hep-ex/0005012.

\bibitem{Salam:2007xv}
G.~P. Salam and G.~Soyez,
\newblock JHEP {\bf 05}, 086 (2007), arXiv:0704.0292.

\bibitem{Stirling:1991ds}
W.~J. Stirling,
\newblock J. Phys. {\bf G17}, 1567 (1991).

\bibitem{Catani:1993hr}
S.~Catani, Y.~L. Dokshitzer, M.~H. Seymour, and B.~R. Webber,
\newblock Nucl. Phys. {\bf B406}, 187 (1993).

\bibitem{Ellis:1993tq}
S.~D. Ellis and D.~E. Soper,
\newblock Phys. Rev. {\bf D48}, 3160 (1993), arXiv:hep-ph/9305266.

\bibitem{Cacciari:2005hq}
M.~Cacciari and G.~P. Salam,
\newblock Phys. Lett. {\bf B641}, 57 (2006), arXiv:hep-ph/0512210.

\bibitem{Cacciari:2008gp}
M.~Cacciari, G.~P. Salam, and G.~Soyez,
\newblock JHEP {\bf 04}, 063 (2008), arXiv:0802.1189.

\bibitem{Weinzierl:2010cw}
S.~Weinzierl,
\newblock Eur. Phys. J. {\bf C71}, 1565 (2011), arXiv:1011.6247.

\end{thebibliography}
\bibliographystyle{/home/stefanw/latex-style/h-physrev5}

\end{document}